# Ordered phases in coupled nonequilibrium systems: static properties

Shauri Chakraborty(1), Sakuntala Chatterjee(1) and Mustansir Barma(2)
*(1) Department of Theoretical Sciences, S.N. Bose National Centre for Basic Sciences, JD Block, Sector 3, Salt Lake, Kolkata - 700106, India*
*(2) TIFR Centre for Interdisciplinary Sciences, Tata Institute of Fundamental Research, Gopanpally, Hyderabad 500107, India.*

We study a coupled driven system in which two species of particles are advected by a fluctuating potential energy landscape. While the particles follow the potential gradient, each species affects the local shape of the landscape in different ways. As a result of this two-way coupling between the landscape and the particles, the system shows interesting new phases, characterized by different sorts of long ranged order in the particles and in the landscape. In all these ordered phases the two particle species phase separate completely from each other, but the underlying landscape may either show complete ordering, with macroscopic regions with distinct average slopes, or may show coexistence of ordered and disordered regions, depending on the differential nature of effect produced by the particle species on the landscape. We discuss several aspects of static properties of these phases in this paper, and we discuss the dynamics of these phases in the sequel.

## I. INTRODUCTION

In a wide variety of physical systems, the constituent microscopic components are driven together to form clusters. As clustering can strongly affect the functioning of the constituents, it is important to understand the factors which influence the nature and degree of clustering, including back-action of the clustering species on the driving field.

Consider an example in a biological setting. Proteins and lipids on the membrane of a living cell are found to cluster, advected by fluctuations of the actin cytoskeleton [1, 2]. A recent model considers these membrane components to be advected passively, in which case the clusters are not stable, and reorganize constantly [3]. However, there is experimental evidence that the membrane components also act back on the actin [4, 5]. The resulting two-way coupling has the potential to affect qualitatively the nature of clustering. This is true in a generic sense: depending on the form and strength of the two-way coupling, clusters may be stable, compact objects or dynamical entities which keep forming and disintegrating on a rapid time scale.

In this paper, we study a coupled nonequilibrium system and investigate in detail the different regimes that emerge as the two-way coupling is varied, with each regime corresponding to a qualitatively different sort of organization. Our system consists of two species of particles moving stochastically on a fluctuating energy landscape [6]. One species is lighter ($L$) while the other is heavier ($H$). The particles tend to minimize their energy by moving along the local potential gradient and also by modifying the landscape around their position to further lower the energy. Thus the $H$ particles preferentially displace the $L$ particles upward while sliding downward along the landscape. Further, each species affects the local landscape dynamics differently.

Particles satisfy a no overlap constraint, with the consequence that the clusters spread out and are macroscopic; they may constitute ordered phases which occupy a finite fraction of the available space. Note that since we are dealing with a nonequilibrium system, the usual strictures of equilibrium statistical mechanics do not apply, and phase transitions between macroscopically different phases can arise, even in one dimension. Indeed it was found that there were several different phases that arise in both one and two dimensions, as the particle-landscape couplings are varied, and the phase diagram was presented in an earlier shorter paper [6]. In the current paper, we discuss the static properties of each phase, while we discuss dynamic properties in the sequel [10]. We present several new results based on analytical calculations and numerical simulations, which help to better understand the origin and properties of the phases reported in [6]. A schematic depiction of typical configurations in each phase is given in Fig. 1. Below we summarize the principal features of each phase, and the new results obtained for each.

**SPS (Strong phase separation)** occurs when the $H$ particles impart a downward push to the surface, and the $L$ particles impart an upward push. This results in complete phase separation between the $H$ and $L$ particles, and between the positive and negative height gradient regions of the landscape as well (Fig. 1a). This is the phase studied in the Lahiri-Ramaswamy model of sedimenting colloidal crystals [7, 8]. The pure domains of positive and negative slope form a macroscopically deep $V$-shaped valley holding the $H$-cluster, while its mirror image /\ holds the $L$-cluster. In [11] a similar phase was reported for the closely related ABC model.

Earlier it was shown that with some conditions on the rates, the steady state measure is given by a Boltzmann factor involving a long-ranged Hamiltonian, if the numbers of $H$ and $L$ particles are equal [8]. In the current paper, this is now generalized to the case of an arbitrary ratio of $H$'s to $L$'s, for which the form of the Hamiltonian is derived. Further, by rescaling rates downward by a factor proportional to the system size, a mean field calculation points to a finite temperature phase transition from a disordered phase to one with long range order. This result is supported by numerical simulations of the model.

**IPS (Infinitesimal current with phase separation)** is obtained when the $H$ particles tend to push the landscape downward, while the $L$ particles do not impart any local bias to the landscape dynamics. In the steady state, the $H$ and $L$ species undergo complete phase separation as in the SPS phase. However, unlike SPS, the landscape is long-range ordered only in the region that holds the $H$-cluster, where it forms a deep valley consisting of macroscopic pure domains of positive and negative slope regions. The remaining part of the landscape beneath the $L$-cluster is not ordered and assumes a rough, parabolic shape (Fig. 1b). Further, in steady state, there is a current of macroscopic tilt (slope) variables through the system with periodic boundary conditions. Representing a positive slope by a particle and a negative slope by a hole, this movement is well described as a SEP (symmetric exclusion process) with input and exit of particles at the two ends of the $L$-region [18, 19]. The value of this current scales inversely with the system size $N$, implying that for large system size, the entire system falls downward at an infinitesimal rate. This accounts for the earlier nomenclature (Infinitesimal fall with phase separation) used for this phase [6].

Some analytic results are obtained for this phase in this paper. First, the Kolmogorov condition [15] for equilibrium is used to demonstrate the breakdown of detailed balance in this case. Further, it is shown that a single $H$ particle in a system of $(N-1)$ particles of type $L$, leads to a non-trivial landscape profile and a current of order $1/N$. Next, the tendency of particles to cluster is demonstrated by considering a system of two $H$ particles and calculating the energy as a function of separation, in the adiabatic limit of vanishingly small rates for particle movement.

Finally, detailed numerical evidence is gathered in support of the description of the landscape in the $L$ region as a SEP with boundary injection.

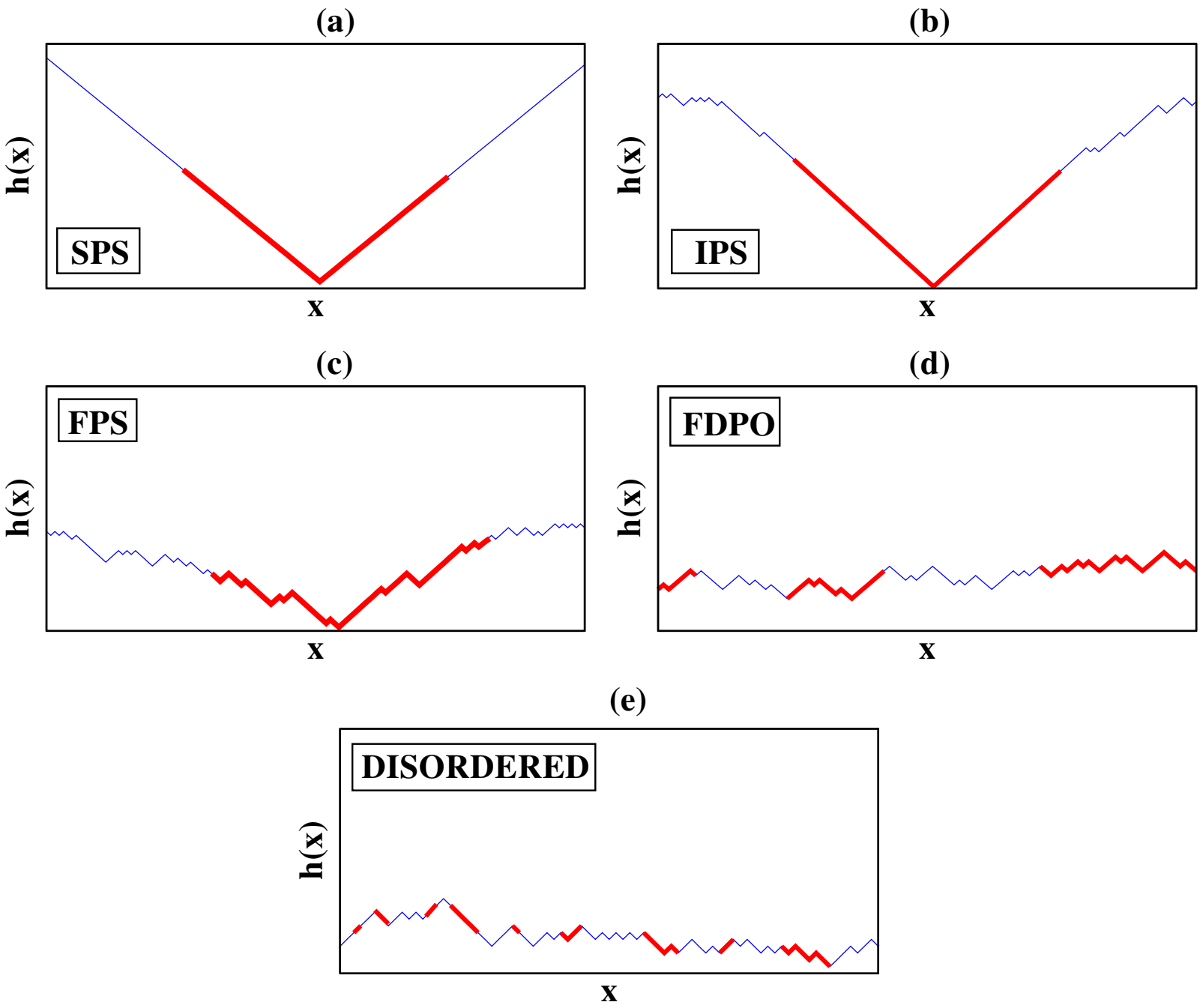

FIG. 1. We show typical configurations of our system in each of the phases. $h(x)$ denotes height of the landscape at a position $x$. The thick (red) lines represent the region occupied by the $H$-particles while remaining parts (blue) are occupied by $L$-particles. (a): A typical configuration in the SPS phase. The clean V-shape of the valley indicates complete phase separation between the upslopes and downslopes in the landscape. (b): A representative configuration in the IPS phase. The landscape occupied by the $H$ particles shows pure domains while the rest has a linear profile with a gradient $1/N$. (c): A typical configuration in the FPS phase. Unlike SPS or IPS phases, the landscape does not have a compact domain of upslope and downslope bonds. Although there is a macroscopic valley, the finite fraction of downslopes (upslopes) present in majority upslope (downslope) domain makes is have a smaller slope. Also, the valley bottom is often rough and more than one minimum can be present. (d): In the FDPO phase, the order in the surface is completely lost. However, the particles congregate into several macroscopic clusters. (e): In the disordered phase, both the particles and the landscape are devoid of any order.

**FPS (Finite current with phase separation)** sets in when both $H$ and $L$ particles push the landscape downwards, but the latter at a lower rate than the former. As in the IPS phase, the $H$ and $L$ species segregate into pure phases and the landscape forms a macroscopic valley holding the $H$ cluster while the part beneath the $L$ cluster is disordered (Fig. 1c). However, unlike the IPS phase, the two arms of the macroscopic valley now have a slope of magnitude less than unity, corresponding to a finite fraction of both tilt species being present in both the arms. The entire system carries a finite current of tilts in the steady state, resulting in a net downward motion with finite velocity. This accounts for the earlier name (fast fall with phase separation) given to this phase [6]. The movement of microscopic tilts in the $L$-region resembles the movement of particles and holes in the well-known ASEP (asymmetric simple exclusion process), with boundary injection [20].

The fact that the steady state tilt current must be uniform across regions, allows us, at the level of mean field theory, to relate the slopes of the arms in the $H$ region to the tilt current in the $L$ region. This value of the slope is shown to be close to that obtained by numerical simulations. Further, the argued-for correspondence of the surface in the $L$ region and the ASEP is tested by numerical simulations. Results conform surprisingly well with the maximal current phase of the ASEP, including for instance the power laws which characterize the density profiles near the edges.

**FDPO (Fluctuation-dominated phase-ordering)** occurs when both $H$ and $L$ particles push the surface down equally (Fig. 1d). Then the surface has a uniform downward drift and evolves as per Kardar-Parisi-Zhang (KPZ) dynamics [16], independent of the particle configuration. The movement of $H$ particles is exactly that of passive sliders, which move down the surface slope while the surface evolves autonomously [9]. The passive slider problem has been well studied in the past, so we do not pursue it here. Particles form macroscopic clusters, with long-range order but macroscopic clusters undergo constant reorganization on the disordered landscape [9]. The spatial correlation function for the particles shows a cusp singularity for small values of the separation scaled by system size $N$.

In the **Disordered phase**, depicted in Fig. 1e, both the $H$ and $L$ particles push the surface down, but the latter at a larger rate than the former. In this case, $H$ particles slide towards local valleys but then such valleys do not remain stable as $L$-rich regions fall faster. With rapid reconfiguring, neither the particles nor the landscape show long-range order, and the phase is disordered.

Similar phases also occur in two dimensions. In the ordered phases, the landscape organizes itself to form a valley with a diamond-shaped cross section, which supports the $H$-cluster. However, there is an interesting finite size effect which gives rise to a different topology of the landscape for smaller systems. Instead of a deep valley with a single minimum, the landscape develops a line of minima and assumes the shape of a trench. Using a scaling argument we show that in the thermodynamic limit, such configurations are energetically unfavorable in comparison to the diamond-shaped single valley.

We limit ourselves to discussion of static properties in this paper and present results on dynamics in a sequel [10]. The rest of the paper is organized as follows. In the next section, we describe our model. In section III we first present the phase diagram and then describe the static properties of the SPS, IPS and FPS ordered phases. In section IV we present the results for two dimensions. Our conclusions are presented in the last section.

## II. THE *LH* MODEL

Our model describes two sets of hard-core particles sliding under gravity on a fluctuating landscape. The local dynamics of the particles and the landscape are coupled such that the particles (a) tend to minimize their energies by moving along the local height gradient of the landscape and (b) modify the landscape around their positions so as to minimize the energy further.

It is useful to think of the set-up as a system of lighter ($L$) and heavier ($H$) particles moving under gravity on a fluctuating surface. In a more generic language, the model consists of two coupled conserved fields — the particle density and the height gradient of the landscape, evolving under a mutual interaction that aids minimization of energy of the whole system.

In one dimension, the LH model consists of a periodic chain lattice of length $N$ where the particles reside on the sites, while the bonds, representing discrete surface elements, can have two possible orientations with slopes $\tau_{i+1/2} = \pm 1$. The $H$ and $L$ particles at neighboring sites may interchange locations preferentially if the local tilt of the surface favors a downward move for $H$. The particles interact via hard-core exclusion and hence a site can be occupied by at most one $H$ or $L$ - particle. The parts of the landscape rich in $H$ particles get pushed down at a higher rate than the parts occupied by $L$ particles. The symbols '/' and '\' indicate up-slope ($\tau_{i+1/2} = 1$) and down-slope ($\tau_{i+1/2} = -1$) bonds, respectively. The update rules for the particles are:

$$
\begin{aligned}
W(H\backslash L \to L\backslash H) &= D + a \\
W(L\backslash H \to H\backslash L) &= D - a \\
W(H/L \to L/H) &= D - a \\
W(L/H \to H/L) &= D + a
\end{aligned}
\tag{1}
$$

where $W$ denotes the probability per unit time for each event to occur. The above dynamics conserves the total number of $H$ and $L$ particles on the lattice. Similarly, the total number of up and down-slopes are also conserved in the dynamics of the landscape:

$$
\begin{aligned}
W(/H\backslash \to \backslash H/) &= E + b \\
W(\backslash H/ \to /H\backslash) &= E - b \\
W(/L\backslash \to \backslash L/) &= E - b' \\
W(\backslash L/ \to /L\backslash) &= E + b'
\end{aligned}
\tag{2}
$$

We consider an untilted surface such that $\sum_i \tau_{i+1/2} = 0$. Note that interchanging $b$ and $b'$ is equivalent to interchange of the $H$ and $L$ particles.

This model was defined in [8] in the context of sedimenting colloidal crystals with the two species representing gradients of the longitudinal and shear strains, but only the case $b = b' > 0$ was studied there.

The model may be generalized straightforwardly to describe $H$ and $L$ particles on a two-dimensional surface. This is described in Section IV below.

## III. RESULTS: PHASE DIAGRAM AND DIFFERENT ORDERED PHASES IN THE SYSTEM

As we vary the transition rates in Eqs. 1 and 2 we encounter different phases. We restrict this variation to the regime in which the parameters $a$ and $b$ in these equations remain positive. In other words, the $H$ particles always show a tendency to slide downhill and push the landscape downward. The ratio

$$R = \frac{E-b}{E+b} \tag{3}$$

then always remains bounded between 0 and 1. The constant $b'$, on the other hand, can be positive, negative, or zero and the ratio

$$R' = \frac{E+b'}{E-b'} \tag{4}$$

can take any value between 0 and $\infty$. The resulting differences in the action of the $L$ particles have macroscopic consequences, and result in different phases.

For $b' > 0$, or $1 < R' < \infty$ the part of the surface containing $L$ particles has a bias to move upward; the macroscopic consequence is that an SPS phase is obtained. For $b' = 0$, or $R' = 1$, the landscape beneath the $L$ particles has unbiased local fluctuations, and on the macroscopic scale an IPS phase results. For $-b < b' < 0$, or $R' < R$, the $L$ particles push the landscape downward, but with a rate smaller than the $H$ particles do, and we have an FPS phase. The limit $b' = -b$, or $R = R'$ corresponds to the case when $H$ and $L$ particles behave identically and we have an FDPO state, characterized by weak phase ordering among the particles and a disordered landscape. For $-b > b'$ the $L$ particles push the surface downward at a larger rate than $H$ particles and in this case neither the landscape nor the particles show any ordering and the system is in a homogeneous or disordered phase. Fig. 2 shows the phase diagram of the system in the $R - R'$ plane. A brief description of the different phases is given in Table III.

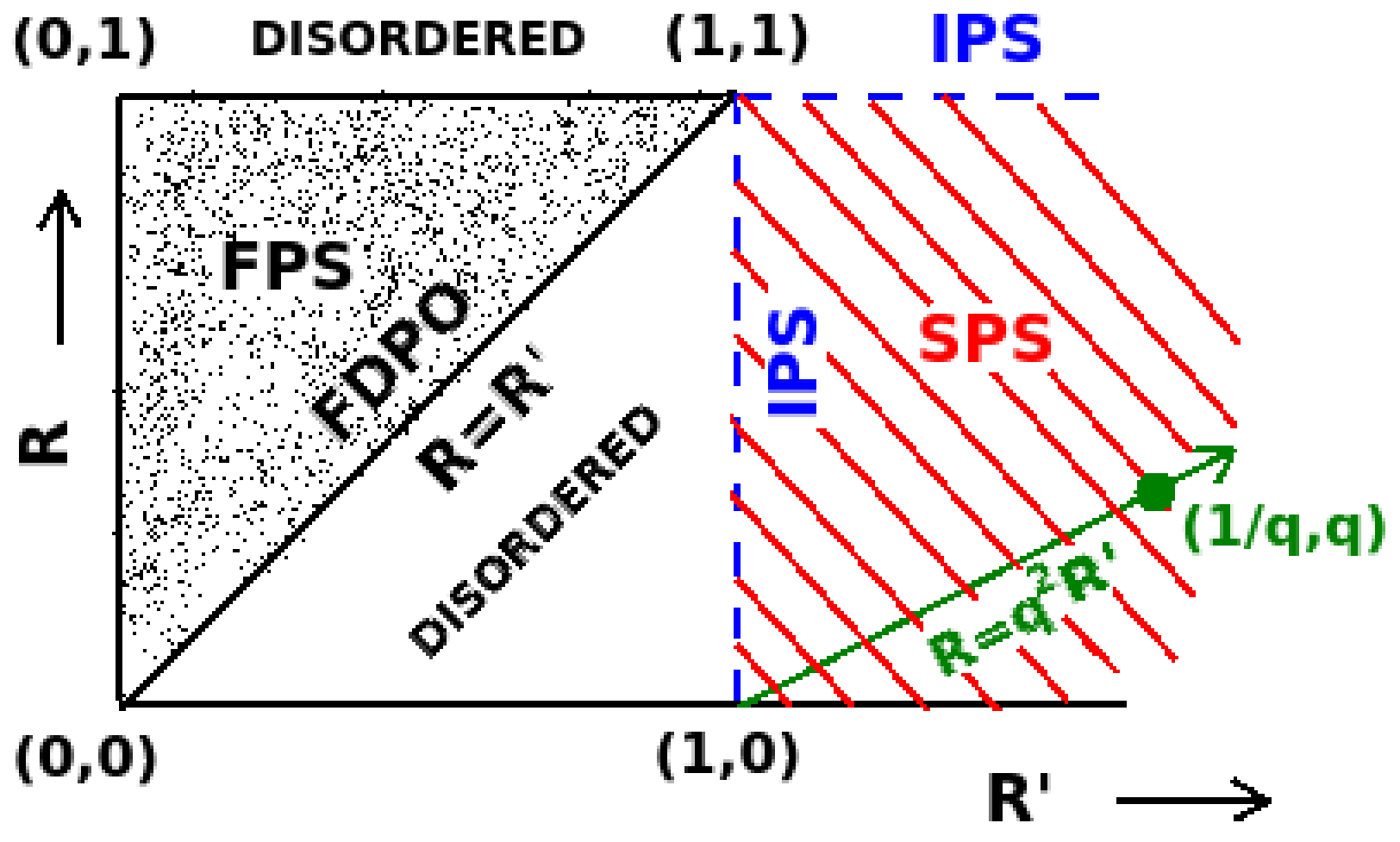

FIG. 2. Phase diagram in the $R - R'$ plane, where $R = \dfrac{E-b}{E+b}$ and $R' = \dfrac{E+b'}{E-b'}$ with model parameters $E$, $b$ and $b'$ defined in Eq. 2 as part of model description. Here, $R > 1$ ($R' < 1$) indicates a downward bias imparted by $H$ ($L$) particles on the landscape, and $R' > 1$ indicates that $L$ particles push the landscape upward. For $1 < R' < \infty$, one has the SPS phase. In this regime, detailed balance in satisfied in the system on the straight line $R = q^2 R'$ where, $q = (D-a)/(D+a)$. The LR model is shown by a solid circle on this line. The dashed lines shown in the diagram correspond to the IPS phase. The dotted region corresponds to the FPS phase ($R' < R < 1$), while the white region corresponds to the disordered phase ($R < R' < 1$). Disordered phase is also seen when $R' < 1$ and $R = 1$. For $R = R' < 1$ FDPO phase is observed.

| Phase | Condition | Particles | Landscape | Downward velocity |
|---|---|---|---|---|
| (a)SPS | $b' > 0$ | Single, compact, macroscopic $H$ and $L$ clusters | Complete phase separation of up-slope and down-slope bonds | $\sim \exp(-\alpha N)$ |
| (b)IPS | $b' = 0$ | | Deep valley beneath $H$ cluster and disordered slopes with gradient $\sim 1/N$ below $L$ cluster | $\sim 1/N$ |
| (c)FPS | $-b < b' < 0$ | | Partial phase separation of slopes beneath $H$ cluster and disordered beneath $L$ cluster | Finite |
| (d)FDPO | $b' = -b$ | Several compact macroscopic clusters of fluctuating lengths | Disordered | Finite |
| (e) Disordered | $-b > b'$ | No macroscopic clusters | Disordered | Finite |

As seen from the discussion in the previous paragraph, the $b' = -b$ or $R = R'$ line acts as the boundary between ordered and disordered phases. This can be also be seen directly from a linear stability analysis of the corresponding continuum theory, describing the system as a coupled time evolution of two conserved fields, the density field of the particles and the tilt field (or height gradient) of the landscape. One can write down the continuity equations in terms of the particle current and the tilt current. Denoting the coarse-grained particle density as $\rho(x,t)$ and landscape height gradient as $m(x,t)$, the corresponding currents within mean-field approximation are given by

$$J_\rho = 2a\rho(x,t)[1-\rho(x,t)][1-2m(x,t)] \tag{5}$$

$$J_m = m(x,t)[1-m(x,t)][2\rho(x,t)(b+b') - 2b'] \tag{6}$$

In the disordered phase, we can use hydrodynamic expansion of $\rho(x,t)$ and $m(x,t)$ about the homogeneous state and retain only linear terms in $\delta\rho(x,t) = \rho(x,t) - \rho_0$ and $\delta m(x,t) = m(x,t) - m_0$ in the expression for $J_\rho$ and $J_m$. Here, $\rho_0$ is the average density of the $H$ particles and $m_0$ is the average value of the slope. Then the continuity equations for the density and slope lead to

$$\partial_t \begin{pmatrix} \delta\rho \\ \delta m \end{pmatrix} = \begin{pmatrix} 0 & -4a\rho_0(1-\rho_0) \\ (b+b')/2 & 0 \end{pmatrix} \partial_x \begin{pmatrix} \delta\rho \\ \delta m \end{pmatrix} \tag{7}$$

We have used the fact that the overall tilt of the surface is zero. The eigenvalues of the Jacobian matrix are $\lambda_\pm = \pm\sqrt{-2a\rho_0(1-\rho_0)(b+b')}$. For $b < -b'$ these eigenvalues are real and the system remains homogeneous. But for $b > -b'$, these eigenvalues are imaginary, which means linear instability grows with time and takes the system away from the homogeneous state and ordered structures are formed. This shows that the line $b = -b'$ indeed is the boundary between ordered and disordered phases in the phase diagram. In the remaining part of this section, we discuss several aspects of the different ordered phases in detail.

## A. SPS phase ($b' > 0$)

In the striped part of the phase diagram ($R' > 1$), the model is identical to that considered by Lahiri and Ramaswamy [7, 8] in the context of sedimenting colloidal crystals. In this regime, the system exhibits SPS. While the $H$'s push the landscape down, the $L$'s tend to push it upwards. Both particle species and the upslope and downslope bonds of the landscape undergo complete phase separation into a macroscopic valley and a hill that holds the $H$ and the $L$ cluster respectively, as shown in Fig. 1(a).

### *1. Detailed balance in SPS*

In [8] it was shown that if the surface is untilted and the density $\rho$ of the $H$ particles is 1/2, the condition of detailed balance holds with respect to a Hamiltonian $\mathcal{H}$ with long-ranged interactions, provided that the rates obey a certain condition. We show below that detailed balance can be recovered with a generalized form of $\mathcal{H}$ for an arbitrary density $\rho$, with a $\rho$-dependent condition on the rates. In this case, detailed balance holds and the steady state measure of the system is given by $\sim \exp(-\beta\mathcal{H})$ with the following Hamiltonian:

$$\mathcal{H} = \sum_{i=1}^{L} (n_i - \lambda) h_i. \tag{8}$$

Here $h_i$ is the height of the $i$-th site defined as $h_i = \sum_{j=1}^{i-1} \tau_{j+1/2}$ and $n_i$ is the occupancy of the $i$-th site which takes the value 1 or 0, according as the site is occupied by an $H$ or an $L$ particle. The parameter $\lambda$ lies in the range between 0 and 1; the deviation of its value from 1/2 characterizes he degree to which $L-H$ interchange symmetry is broken. It is sufficient to consider the range $0 < \lambda \leq 1/2$ since $\mathcal{H}$ remains invariant under $\lambda \to 1-\lambda$ and $n_i \to 1-n_i$. Figure 3 shows the rates of the allowed microscopic moves along with the change of energy they entail. It is easy to see from this figure that the condition of detailed balance is satisfied for the following choice of rates,

$$\frac{D-a}{D+a} = q, \qquad \frac{E-b}{E+b} = q^{2-2\lambda}, \qquad \frac{E-b'}{E+b'} = q^{2\lambda} \tag{9}$$

where $q = e^{-\beta}$. Note that the system is defined on a ring and hence it is translationally invariant.

The height $h_i$, as defined above, is measured from the first site. If the sites are relabelled such that the site $k$ with height $h_k = \delta$ is the new origin, then the height of all the sites are changed as $h'_i = h_i - \delta$. To ensure that the total energy of the configuration does not change as a result of this relabelling, one must have $\lambda = \rho$, where $\rho$ is the total density of $H$ particles. Thus, in the $\lambda - \rho$ plane, it is only along the locus $\lambda = \rho$ that detailed balance holds with $\mathcal{H}$ given by Eq. 8. Note that $\lambda = \rho = 1/2$ corresponds to the case considered in [8] for the Lahiri-Ramaswamy model.

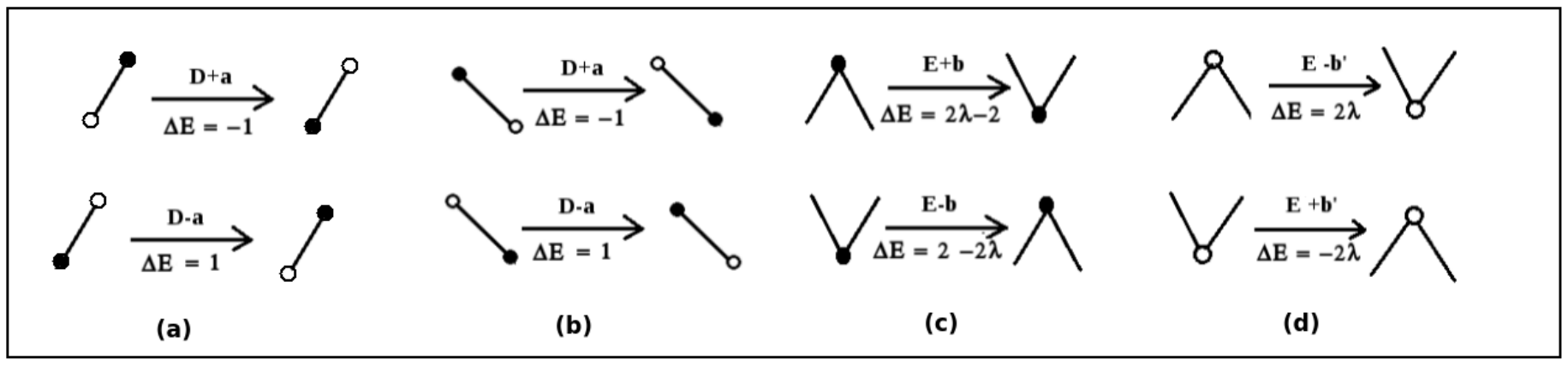

FIG. 3. Schematic representation of different transitions that are allowed to occur in the system. $\Delta E'$s denote the energy costs involved in the transitions, as per the Hamiltonian in Eq. 8. Solid (empty) circles represent $H$ ($L$) particles.

### *2. Rescaled temperature and phase transitions in SPS*

Although the Hamiltonian in Eq. 8 is defined in terms of local height and local occupancy, the definition of the height field generates long-ranged interactions between $n_i$ and $\tau_{j+1/2}$ in the Hamiltonian. This gives rise to a super-extensive energy that scales as $N^2$ which at any finite temperature always wins over the extensive entropy term. In other words, as follows from Eq. 9, for any non-zero $\beta$, or equivalently, any $q < 1$, the $H$-rich phase has a vanishing fraction of $L$ particles, and vice versa. We refer to such phases as 'compact' as they exclude islands of the other species. The name 'strong phase separation' actually refers to this particular aspect of this phase [7, 8]. However, if the parameter $\beta$ is rescaled by system size $N$, *i.e.* $\beta \to \beta/N$, then the energy and entropy terms become comparable. Earlier this was demonstrated in the closely related ABC model [11] for which a similar reduction of rates results in an order-disorder transition at a critical $\beta_c$ [12]. In this section, we present a calculation based on mean field theory to estimate this critical point in our $LH$ model.

Let $\rho_i$ denote the probability to find an $H$ particle at site $i$ and $m_{i+1/2}$ denote the probability to find an upslope bond between sites $i$ and $(i+1)$. Using the dynamical rules described in section II, we can write down the time-evolution equations for these probabilities, within mean-field theory, neglecting all correlations:

$$\frac{d\rho_i}{dt} = \rho_{i-1}(1-m_{i-1/2})(1-\rho_i) - \rho_i(1-m_{i+1/2})(1-\rho_{i+1}) + q\rho_{i-1}m_{i-1/2}(1-\rho_i) - q\rho_i m_{i+1/2}(1-\rho_{i+1}) \tag{10}$$

$$\begin{aligned}\frac{dm_{i+1/2}}{dt} &= m_{i-1/2}\rho_i(1-m_{i+1/2}) + q^{2-2\lambda}(1-m_{i+1/2})\rho_{i+1}m_{i+3/2} + (1-m_{i+1/2})(1-\rho_{i+1})m_{i+3/2} \\ &+ q^{2\lambda}m_{i-1/2}(1-\rho_i)(1-m_{i+1/2}) - m_{i+1/2}\rho_{i+1}(1-m_{i+3/2}) - q^{2-2\lambda}(1-m_{i-1/2})\rho_i m_{i+1/2} \\ &- (1-m_{i-1/2})(1-\rho_i)m_{i+1/2} - q^{2\lambda}m_{i+1/2}(1-\rho_{i+1})(1-m_{i+3/2})\end{aligned} \tag{11}$$

Assuming slow spatial variation of $\rho$ and $m$ fields, we can take the continuum limit and write $\rho_i = \rho(x)$ and obtain the following expansion:

$$\rho_{i\pm1} = \rho(x) \pm \frac{1}{N}\frac{\partial\rho(x)}{\partial x} + \frac{1}{2N^2}\frac{\partial^2\rho(x)}{\partial x^2} + ... \tag{12}$$

Similarly,

$$\begin{aligned}m_{i+3/2} &= m(x) + \frac{1}{N}\frac{\partial m(x)}{\partial x} + \frac{1}{2N^2}\frac{\partial^2 m(x)}{\partial x^2} + ... \\ m_{i-1/2} &= m(x) - \frac{1}{N}\frac{\partial m(x)}{\partial x} + \frac{1}{2N^2}\frac{\partial^2 m(x)}{\partial x^2} + ...\end{aligned} \tag{13}$$

Next, we write $q = e^{-\beta/N} = 1 - \frac{\beta}{N} + \frac{\beta^2}{2N^2} + ...$, in which the parameter $\beta$ has been explicitly scaled by the system size. The time-evolution equations 10 then become

$$\frac{\partial\rho}{\partial t'} = \frac{\partial^2\rho}{\partial x^2} + 2\beta\rho(1-\rho)\frac{\partial m}{\partial x} + \beta(2m-1)(1-2\rho)\frac{\partial\rho}{\partial x} \tag{14}$$

$$\frac{\partial m}{\partial t'} = \frac{\partial^2 m}{\partial x^2} - 2\beta\frac{\partial}{\partial x}[\rho m(1-m)] + 2\beta\lambda\frac{\partial}{\partial x}[m(1-m)] \tag{15}$$

where $t' = t/N^2$ is the rescaled time.

In the stationary state, the time-derivatives on the left hand sides of the Eqs. 14 and 15 vanish. Recalling that the overall density of upslope bonds in the system is 1/2 and periodic boundary condition requires the density of $H$ particles to be equal to $\lambda$, we linearize $m(x) = 1/2 + \delta m(x)$ and $\rho(x) = \lambda + \delta\rho(x)$, in the stationary state to obtain

$$\frac{\partial^2}{\delta x^2}\delta m - \frac{\beta}{2}\frac{\partial}{\partial x}\delta\rho = 0 \tag{16}$$

$$\frac{\partial^2}{\delta x^2}\delta\rho + 2\beta\lambda(1-\lambda)\frac{\partial}{\partial x}\delta m = 0 \tag{17}$$

Making the Fourier expansions $\delta m(x) = \sum_n a_n \exp(2\pi inx/N)$ and $\delta\rho(x) = \sum_n b_n \exp(2\pi inx/N)$ we find from Eqs. 16 and 17 that

$$i2\pi n a_n = \frac{\beta}{2} b_n \tag{18}$$

and

$$i2\pi n b_n = -2\beta\lambda(1-\lambda)a_n. \tag{19}$$

To obtain non-zero solutions for $a_n$ and $b_n$ we must have

$$\beta = \frac{2\pi n}{\sqrt{\lambda(1-\lambda)}} \tag{20}$$

which has the minimum value $\beta_c = \dfrac{2\pi}{\sqrt{\lambda(1-\lambda)}}$ for $n = 1$. For any $\beta$ smaller than this value no non-zero $a_n$ and $b_n$ can be found and $\rho(x)$ and $m(x)$ only allow uniform solutions, corresponding to a disordered state ($a_0$ and $b_0$ non-zero). Thus $\beta_c$ gives the critical point for the order-disorder transition in the system.

To verify this in simulations, we define the order parameters as

$$s_\rho = \frac{1}{N}\sum_{i=1}^{N} n_i n_{i+1} - \lambda^2 \tag{21}$$

$$s_m = \frac{1}{4N}\sum_{i=1}^{N}(1+\tau_{i+1/2})(1+\tau_{i+3/2}) - \frac{1}{4} \tag{22}$$

which characterize the order in the particles and the landscape, respectively. Our simulations show that for small values of $\beta$, the average values $\langle s_\rho\rangle$ and $\langle s_m\rangle$ are zero, indicating a disordered phase. As $\beta$ increases, the system goes into an ordered phase with finite values of $\langle s_\rho\rangle$ and $\langle s_m\rangle$. To calculate the critical $\beta$, at which the transition takes place, we plot the second order Binder cumulant $f_\alpha = 1 - \langle s_\alpha^2\rangle/\langle s_\alpha\rangle^2$ as a function of $\beta$ (see Fig. 4), where $\alpha = \rho, m$ for different system sizes. At the critical point $\beta_c$, the value of $f_\alpha$ in expected to be universal, which means the curves for different $N$ values should coincide at $\beta_c$ [12, 13]. In Fig. 4 we present data for $\lambda = 1/5$ for which we expect $\beta_c = 5\pi \simeq 15.708$. From our simulation data we find $\beta_c \simeq 15.65$ and 15.76 for $\alpha = \rho$ and $\alpha = m$, respectively. These values are close to the theoretical prediction.

### B. IPS phase ($b' = 0$)

The IPS phase is obtained along the dashed lines of the phase diagram in Fig. 2, where $b'$ vanishes. Along the vertical dashed line, we have $R' = 1$ and from Eq. 9 it follows that $\lambda = 0$. In this case, the local fluctuations in the surface occupied by $L$ particles are of the symmetric Edwards-Wilkinson type [14], while the $H$ particles continue to push the surface down. In our derivation of the detailed balance condition in section III A 1 above, we have shown that $\lambda = \rho$ has to be satisfied for a periodic system. In the IPS phase this condition is violated for all finite $\rho$. As a result, detailed balance breaks down. It is instructive to use the Kolmogorov loop condition [15] to explicitly show the lack of detailed balance, as illustrated below.

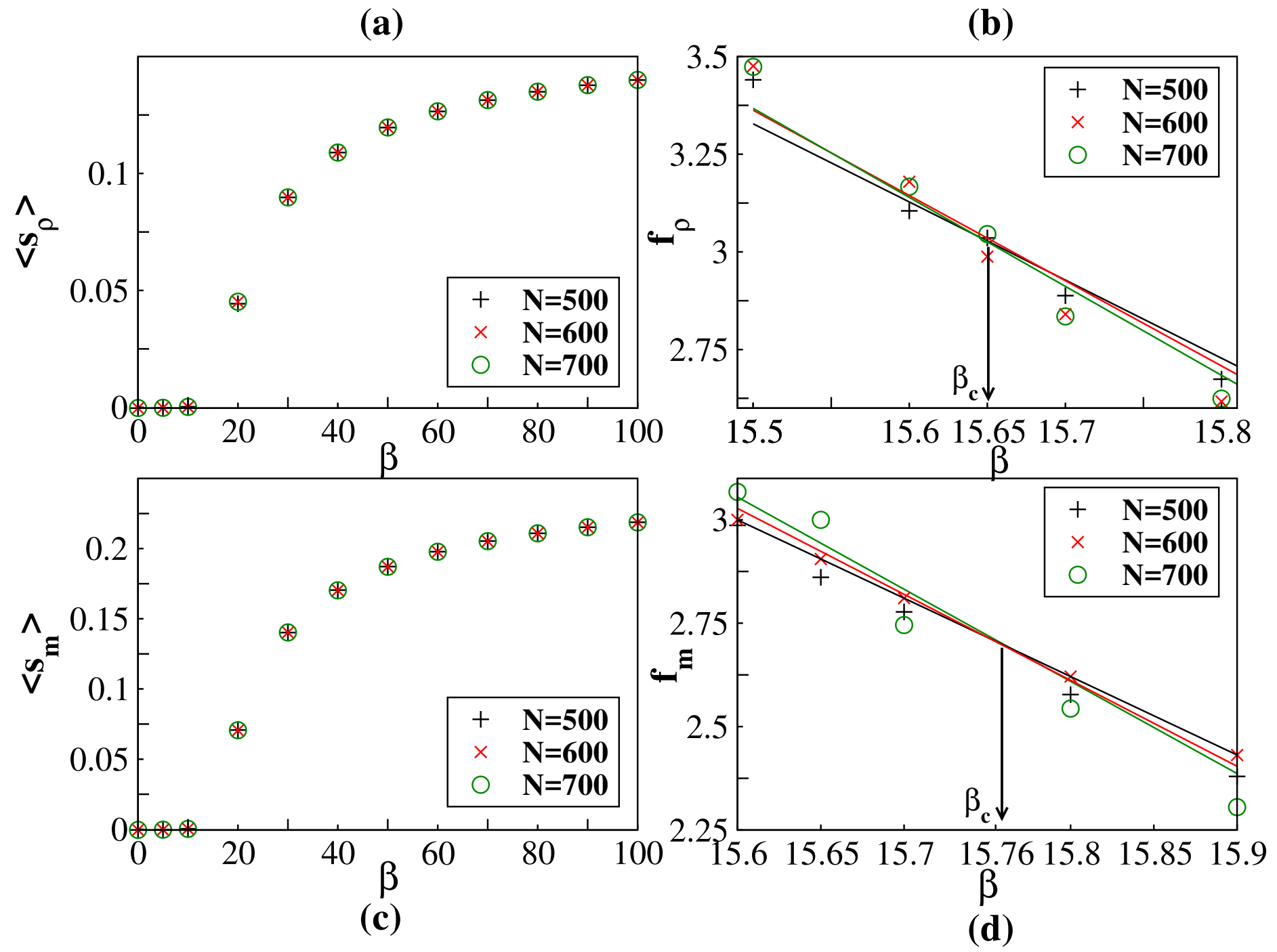

FIG. 4. Temperature variation of $s_\rho$ and $s_m$ for three different values of $N$ [plots (a) and (c)]. (b) and (d) show the cumulants $f_\rho$ and $f_m$ for three $N$ values. We have used $\lambda = 0.2$ here. We obtain the best linear fits to the data points for each value of $N$. From the point of intersection of the straight lines , $\beta_c$ is estimated to be $15.65 \pm 0.001$ for $f_\rho$ and $15.76 \pm 0.005$ for $f_m$ which is close to theoretical prediction $5\pi$. The data shown here have been averaged over at least $10^8$ histories.

### 1. *Breakdown of detailed balance in IPS*

The Kolmogorov loop condition [15] states that the necessary and sufficient condition for a system to satisfy detailed balance is that for every closed loop in configuration space, we must have

$$Q = \frac{W(1 \to 2)W(2 \to 3)...W(K \to 1)}{W(2 \to 1)W(3 \to 2)...W(1 \to K)} = 1 \tag{23}$$

where $W(i \to j)$ denotes the transition rate from configuration $C_i$ to $C_j$. To show that detailed balance is violated, it suffices to find a single loop in configuration space for which the above condition is not satisfied. In Fig. 5 we explicitly show this for a set of local configurations. Since each configuration is specified by the particle occupancy at the lattice sites, and slope of the lattice bonds, the first and the last configurations in the sequence presented in Fig. 5 are identical and hence this sequence forms a closed loop in the configuration space. According to our dynamical rules, $W(4 \to 5) = E + b$ and $W(5 \to 4) = E - b$. All other rates $W(i \to i+1)$ are the same as the reverse rate $W(i+1 \to i)$, since in the IPS phase $b' = 0$. The ratio $Q$ then becomes $Q = \frac{D+a}{D-a} \neq 1$, which proves violation of detailed balance.

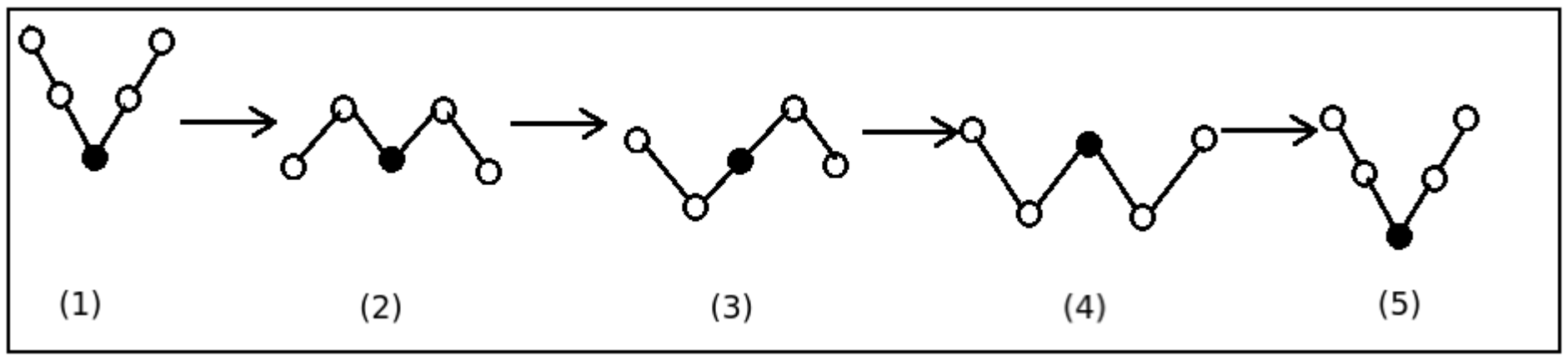

FIG. 5. Breakdown of Kolmogorov loop condition in the IPS phase. Starting from the first configuration, the system passes through a sequence of configurations and comes back to the starting configuration again, but the ratio $Q = \frac{W(1\to2)W(2\to3)W(3\to4)W(4\to5)}{W(2\to1)W(3\to2)W(4\to3)W(5\to4)} \neq 1$. This shows that the system does not obey detailed balance.

### 2. A single $H$ particle in IPS: height profile of the landscape

In order to understand the nature of the IPS phase, first let us consider the case of a single $H$ particle with $(N-1)$ lattice sites occupied by $L$ particles. According to the dynamical rules, the local height fluctuations at these $(N-1)$ sites are symmetric, of Edwards-Wilkinson type [14], and only at the site containing the $H$ particle the height fluctuation is asymmetric, of Kardar-Parisi-Zhang type [16]. Obviously, this asymmetry drives the system out of equilibrium and there is a non-zero current in steady state, which gives rise to a downward velocity of the surface. Since the local fluctuations are symmetric almost everywhere in the system, to support this downward drift, a gradient is generated in the density of upslope and downslope bonds of the surface. We calculate this gradient within mean-field theory below.

Let us consider a site at a distance $k$ from the position of the single $H$ particle in the system, and let $S^+(k,N)$ be the probability to find an upslope bond between this site and its right neighbor. Similarly, let $S^+(k-1,N)$ be the probability to find an upslope bond between the site and its left neighbor. Within mean-field theory, the site under consideration will be at the top of a local hill with probability $S^+(k-1,N)[1-S^+(k,N)]$. From this local configuration, the height of the site can decrease with rate $E$, when the local hill flips to a valley (see Eq. 2). Likewise, the probability that the site is at the bottom of a local valley is given by $[1-S^+(k-1,N)]S^+(k,N)$ and from here its height can increase with the same rate $E$. The downward velocity of the surface at this position is then $E[S^+(k-1,N)\{1-S^+(k,N)\} - \{1-S^+(k-1,N)\}S^+(k,N)]$.

In the steady state, this velocity must be the same everywhere in the system and hence independent of $k$. In other words, $[S^+(k-1,N) - S^+(k,N)] = C$, which is a constant. Moreover, for an untilted surface, $\sum_k S^+(k,N) = N/2$. These two relations together imply that $S^+(k,N)$ decreases linearly with $k$ with a gradient $\sim 1/N$, for large $N$. The results of numerical simulations verify this (Fig. 6).

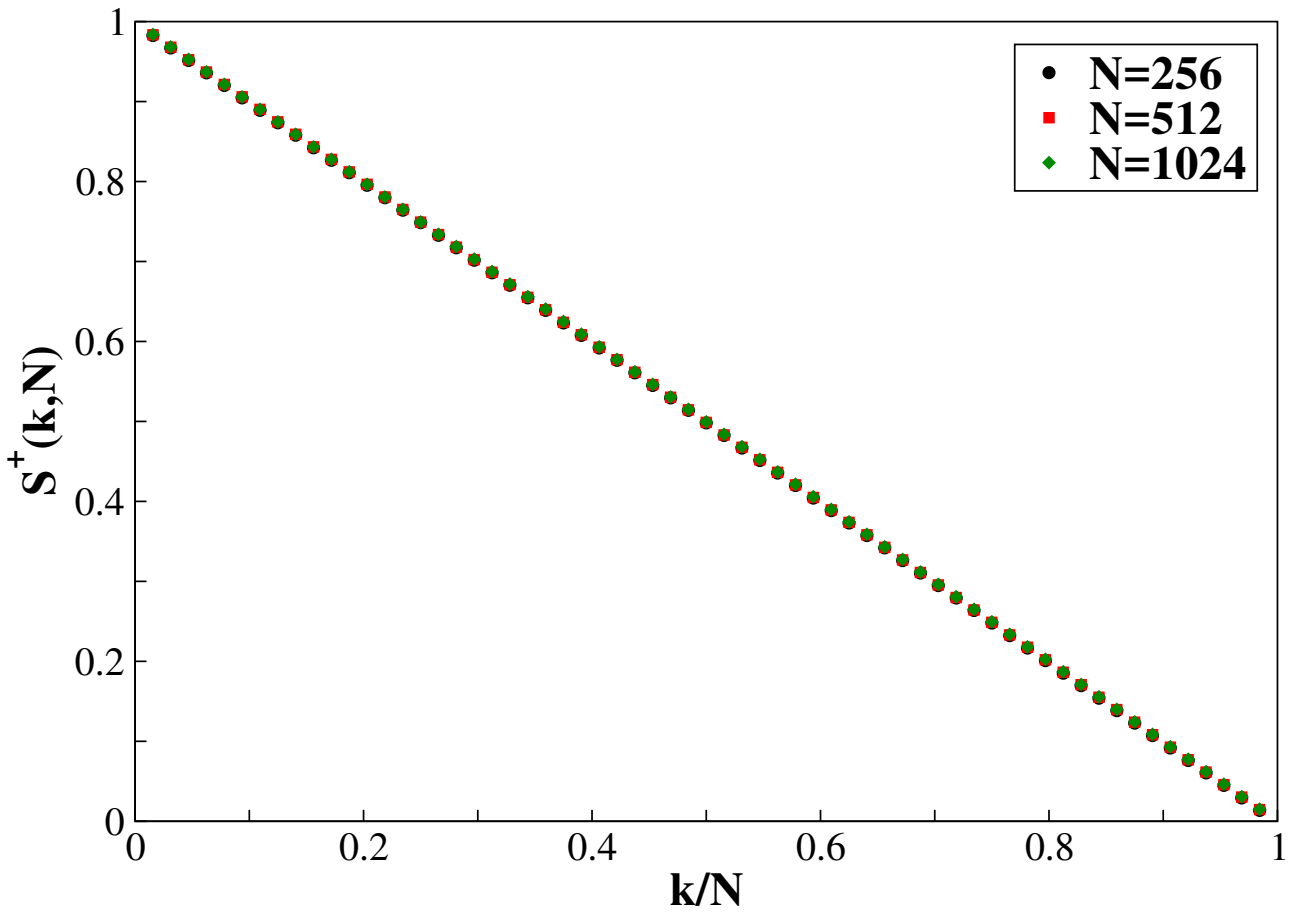

FIG. 6. We measure the density of upslopes $S^+(k,N)$ as a function of the scaled distance $k/N$, where $k$ is the distance measured from the position of the single $H$ particle in a system of size $N$. The profile decreases linearly with a gradient $\sim 1/N$ as predicted by our mean field analysis. The data shown here have been averaged over at least $10^4$ initial histories.

The average height of the landscape at a distance $i$ from the $H$ particle is $\langle h_i \rangle = \sum_{k=1}^{i}[2S^+(k,N)-1]$. Thus in presence of a single $H$ particle the height profile of the landscape has a parabolic shape with the $H$ particle at the minimum height. Our calculations remain valid even when many $H$ particles are present, since these particles form a compact cluster in IPS phase and can be treated as a single entity. $\langle h_i \rangle$ in that case represents the average height of the landscape at a distance $i$ measured from the edge of the $H$-cluster. Thus, our calculation explains the shape of the landscape in the $L$-region of the IPS phase, shown in Fig. 1b.

### 3. Clustering tendency of $H$ particles in the IPS phase in the adiabatic limit

In this section, we present a simple calculation to explain why the $H$ particles tend to form a compact cluster in the IPS phase.

First let us consider a large system in the continuum limit with a finite $M$ number of $H$ particles in it. Let $x_1, x_2, ..., x_M$ be the positions of these particles. Apart from these positions, the local height fluctuations of the surface are symmetric and of Edwards-Wilkinson type, while at the positions $x_i$'s the height fluctuations are biased. This is captured by the equation

$$\partial_t h(x,t) = D\partial_x^2 h(x,t) + \eta(x,t) + j_0 \sum_{n=1}^{M} \delta(x - x_n) \tag{24}$$

where $j_0$ represents the bias imparted by the $H$ particles, $\eta(x,t)$ is the white noise and $D$ the surface diffusivity. To find the mean profile $\overline{h}(x,t)$, we average over the noise and obtain

$$\partial_t \overline{h}(x,t) = D\partial_x^2 \overline{h}(x,t) + j_0 \sum_{n=1}^{M} \delta(x - x_n). \tag{25}$$

Let us first consider $M = 1$, when there is a single $H$ particle present in the system at the position $x_1$. To solve the above equation, we make an adiabatic assumption based on the separation of time-scales. Suppose the $H$ particle is so heavy such that it hardly moves during the time the height fluctuations of the surface are taking place. In this limit, we can treat $x_1$ as the position of a quenched defect and without any loss of generality put $x_1 = 0$. The above equation can then be solved using Green's function method [17]. Starting from a flat height profile at $t = 0$, we can write the height profile at time $t$ as

$$\overline{h}(x,t) = j_0 \int_0^t \frac{1}{(4\pi D)^{1/2}} \frac{e^{-x^2/4Ds}}{s^{1/2}} ds = \frac{j_0}{4D\pi^{1/2}} |x| \Gamma\left(-\frac{1}{2}, \frac{x^2}{4Dt}\right) \tag{26}$$

where $\Gamma$ denotes the incomplete Gamma function. For large $t$, we have $x^2/4Dt \ll 1$ and

$$\overline{h}(x,t) \approx \frac{j_0}{\sqrt{\pi}} \left[ 2\sqrt{\frac{t}{D}} - \frac{\sqrt{\pi}}{2D} \frac{|\, x\,|}{D} \right]. \tag{27}$$

Similarly, for $M = 2$, when there are two close-by quenched defects in the system, at $x_1$ and $x_2$, each will generate a height profile around its position. As the system is linear, the resulting height profile is given by

$$\overline{h}(x,t) \approx \frac{j_0}{\sqrt{\pi}} \left[ 2\sqrt{\frac{t}{D}} - \frac{\sqrt{\pi}}{2D} (|\, x - x_1\,| + |\, x - x_2\,|) \right]. \tag{28}$$

For particles of mass $m$ moving under gravity, $j_0$ is negative and the mean gravitational energy $E_g$ associated with the system is

$$E_g = -\frac{mg|j_0|}{\sqrt{\pi}} \left[ 4\sqrt{\frac{t}{D}} - \frac{\sqrt{\pi}}{D} \,|\, x_1 - x_2\,| \right] \tag{29}$$

which is minimum when $|\, x_1 - x_2\,|$ is minimum. This explains why the two $H$ particles tend to cluster together. This argument can be extended immediately to arbitrary $M$, providing insight into the strong clustering tendency of $H$ particles in the IPS phase.

Indeed for a finite density of $H$ particles in the lattice model, where the particles have a finite size, we find a complete phase separation of $H$ and $L$ particles. The upslope and downslope surface bonds lying under the $H$ particle cluster, phase separate to form a deep valley. The domains of all-upslope and all-downslope bonds extend up till the edges of the $H$ particle cluster. Beyond that, in the $L$-phase, the landscape has a parabolic shape with a mean curvature $1/N$, as shown in section III B 2. The fluctuation properties of the surface beneath the $L$-cluster can be explained by mapping this part of the system to an open-chain symmetric exclusion process [18], with the upslope (downslope) bonds being identified with particles (holes); the pure domains of these bonds in the $H$-phase act as reservoirs for the respective species. We elaborate more on this issue in [10] where we discuss the steady state dynamics.

Compact domains in the IPS phase are observed as long as the $H$ particles act as the heavier species, *i.e.* as long as the ratio $q = (D - a)/(D + a) < 1$. As $q$ approaches unity, the domains do not remain as compact and their boundaries become wider. However, this width remains finite and in the thermodynamic limit, sufficiently far away from these boundaries a pure phase is always retrieved. For a finite system size $N$, there exists a critical value $q_c$, when the width of the domain boundaries becomes of the order of the system size and the system becomes disordered. On performing a mean-field calculation similar to that in section III A 2, we find $q_c = \exp(-4\pi/N)$, which we have verified by simulation (data not shown here).

### C. FPS phase ($-b < b' < 0$)

The FPS phase can be observed in the dotted region of the phase diagram (Fig. 2), where $1 > R > R'$, *i.e.* when both the particle species push the landscape down, but the $H$ particles do so at a larger rate than the $L$'s. In this phase, the $H$ and $L$ particles again show complete phase separation and although the landscape forms a single macroscopic valley, neither of the two arms of the valley comprises a compact domain of / or \ bonds, unlike in the SPS or IPS phases. In this section we present numerical and analytical results on the static characterization of this phase.

#### 1. Static correlations in H-region of the landscape in FPS

A typical configuration in the FPS phase is shown in Fig. 1(c). Here, a large valley forms in the landscape that holds the $H$-cluster, but unlike the IPS phase, this valley consists of domains of upslope and downslope bonds which are not compact. Let $m$ be the density of downslope (upslope) bonds in the upslope-(downslope-)rich domain. For a perfectly ordered domain, $m$ takes the value 0 while in the disordered case $m = 1/2$, while $0 < m < 1/2$ indicates a phase separation with the minority species interspersed with the majority species.

It is possible to analytically calculate the value of $m$ in this phase. Within mean-field theory, it follows from the dynamical rules (Eq. 2) that the average velocity of the surface in $H$-region is $2bm(1-m)$. In the steady state, this must be equal to the velocity in the $L$-region. Now, in the $L$-region the surface is disordered, and this part of the surface can be mapped onto an open-chain asymmetric exclusion process in the maximal current phase [6]. The velocity of the surface in this region is then $b'/2$. Matching the two velocities in the $H$ and $L$ regions, we find the following quadratic equation

$$m^2 - m - \frac{b'}{4b} = 0 \tag{30}$$

which can be solved for $m$ for a given $b$ and $b'$. To test Eq. 30, we numerically measure the density of upslope bonds in the $H$-region and find good agreement (see Fig. 7a). Across the valley minimum, the slope shows a transition from the value $m$ to $(1-m)$. The width of the boundary between the upslope-rich and downslope-rich domains scales as $\sqrt{N}$, as shown in Fig. 7b. In [10] this width was related to the motion of the valley bottom within a region of size $\sim \sqrt{N}$ around the centre of mass of $H$-cluster. It should be noted that the system shows rather strong finite size effects and we had to go to relatively large $N$ ($\sim 10^4$) in order to find the saturation value $m$ of $S^+$ away from the domain boundary.

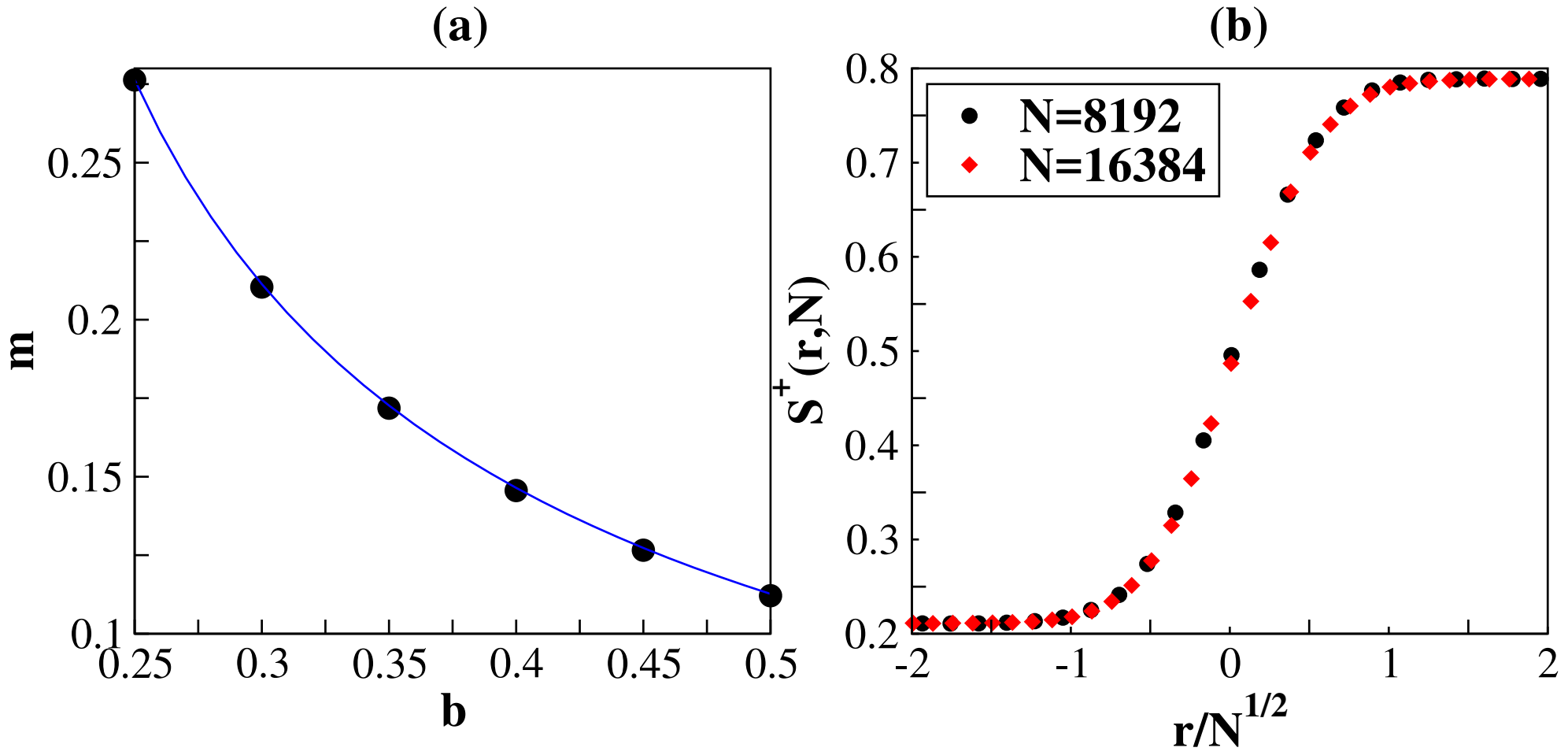

FIG. 7. (a): Away from the valley bottom $S^+(r,N)$ saturates at a value $m$, which depends on $b$ and $b'$. For $b' = -0.3$, we vary $b$ and plot $m$, which matches well with the mean-field result. Discrete points are from simulation data and the continuous line shows mean-field solution of Eq. 30. We have used $N = 1024$ here. (b): $S^+(r,N)$ changes from the value $(1-m)$ to $m$ across the domain boundary of width $\sqrt{N}$.

*2. Disordered landscape in the L-region in FPS*

The behavior of the landscape in the $L$-region is like that of an open system. The ordered domains of the upslope and downslope bonds in the $H$-region act as the reservoirs which are connected to the two ends of the $L$ region. Mapping the upslope (downslope) bonds to particles (holes), we find that the surface in the $L$-region can be mapped onto an open-chain asymmetric exclusion process, which was introduced in [20] and different phases were obtained on changing the reservoir couplings. In the FPS phase, the landscape is disordered in the $L$ region, with $S^+(r,N) = 1/2$. The properties of the landscape in the $L$-region are the same as those observed for a maximal current phase in the open system. Away from the $H-L$ domain boundary, $S^+(r,N)$ decays algebraically to the disordered value $S^+ = 1/2$ with an exponent $1/2$, as shown in Fig. 8.

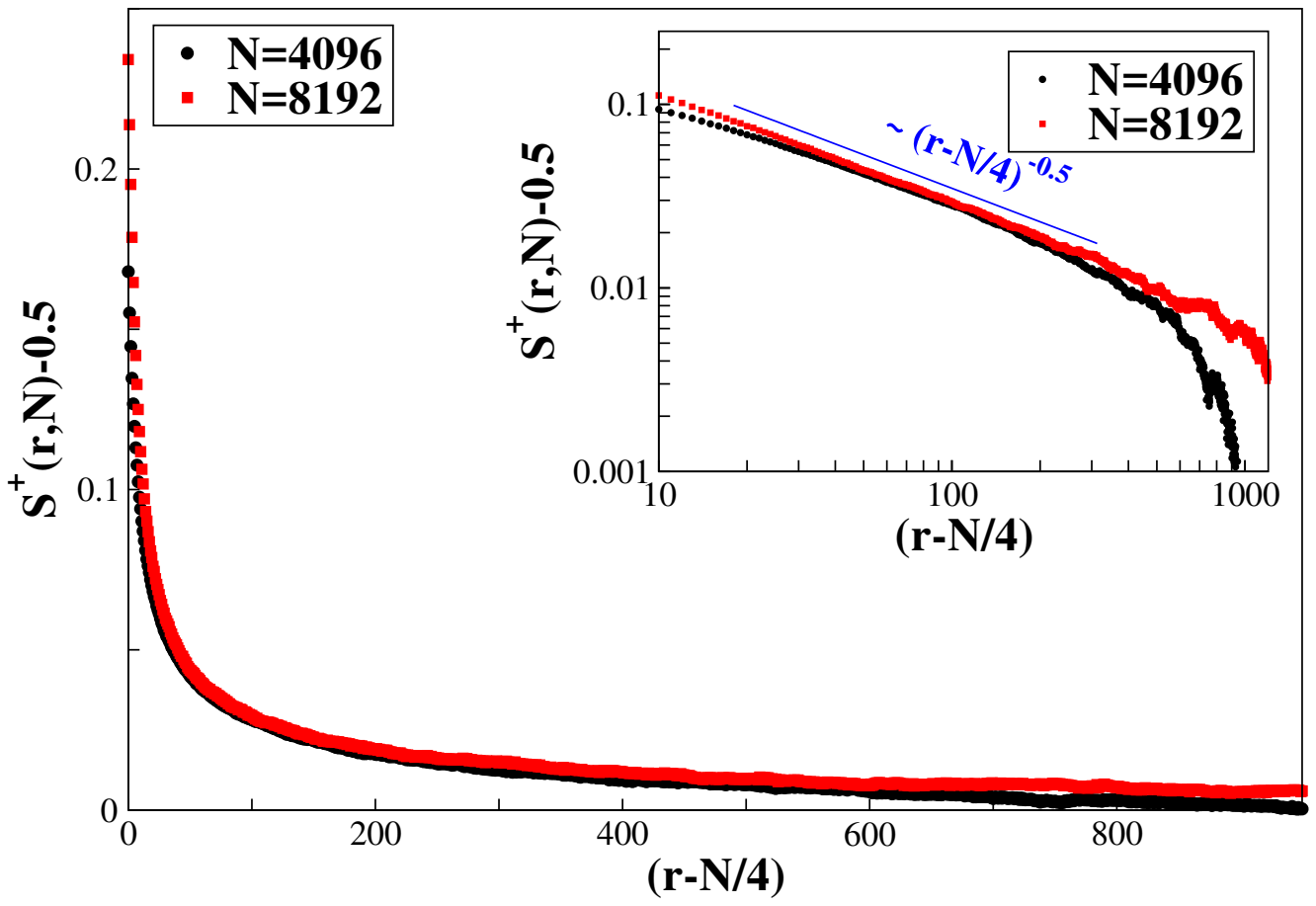

FIG. 8. We plot $S^+(r,N)$ after subtracting the disordered phase value $1/2$ against $(r-N/4)$ for two different system sizes with $b = 0.3, b' = -0.2$. The decay to a disordered phase occurs algebraically with an exponent $1/2$ (inset). The data shown here have been averaged over at least $10^6$ initial configurations.

## D. FDPO phase ($b = -b'$)

The FDPO phase can be observed along the $R = R'$ line in the phase diagram in Fig. 2. In this case, $H$ and $L$ particles push the surface down at exactly the same rate. The transition rates to transform local hills to valleys are therefore identical at every lattice site. In other words, the surface behaves just like the single step model [16], which on large length and time scales is described by the KPZ equation. With periodic boundary conditions, the steady state satisfies product measure [21], *i.e.* the upslope and downslope bonds of the landscape are independently and randomly distributed and the landscape is disordered. This remains true even when both $b$ and $b'$ are zero, and the landscape shows Edwards-Wilkinson type equilibrium fluctuations, since the same product measure holds for a periodic Edwards-Wilkinson surface in one dimension.

Note that in this phase, the coupling between the landscape and the particles is one-way, *i.e.* while the particles continue being affected by the local height gradient of the landscape, with moves shown in Figs. 3a and 3b, the local dynamics of the landscape does not depend on whether there is an $H$ particle or $L$ particle on it. This limit is tantamount to passive scalar advection and was studied in detail in [9]. In this phase, while the landscape remains completely disordered, the $H$ and $L$ particles show clustering accompanied by macroscopic fluctuations, as depicted in Fig. 1d. Owing to strong fluctuations present in the system, these clusters undergo constant reorganization, even in the thermodynamic limit. The two-point correlation function exhibits long-range order, but the Porod law breaks down and the scaled correlation function exhibits a cusp singularity at small argument [9].

## E. Disordered phase ($b' < -b < 0$)

The disordered phase is shown by the unshaded part of the phase diagram in Fig. 2. In this phase neither the landscape nor the particles show long range order. Interesting results on the dynamical correlation functions in this phase will be presented elsewhere [22].

## IV. TWO DIMENSIONS

The phase diagram shown in Fig. 2 remains valid even in two dimensions. We consider an $N \times N$ square lattice and denote the height at the site $(i, j)$ by $h(i, j)$, and ensure that the height difference between the nearest neighbor sites is always maintained at $\pm 1$. Particles still move from one site to a neighboring one and if the height of the destination site is lower, then $H$ particles preferentially displace the $L$ particles, as in Eq. 1. The time-evolution of the two dimensional surface occurs due to transition between the local hills and valleys [23]. A site $(i, j)$ is said to be a local hill (valley) if all its four neighbors with coordinates $(i \pm 1, j)$ and $(i, j \pm 1)$ have height $h(i, j) - 1$ $(h(i, j) + 1)$. The transition rates for hills and valleys occupied by $H$ or $L$ particles remain same as in Eq. 2.

The phases obtained in this case are similar to those in one dimension. In the ordered phases, the $H$ and $L$ particles undergo complete phase separation and the landscape beneath the $H$ cluster orders to form a deep valley. In the SPS phase, the entire landscape is ordered to form a deep valley accommodating the $H$-cluster at the bottom of it. In the IPS phase, the landscape beneath the $L$-region shows a linear variation of height along $x$ or $y$ direction, while in the FPS phase it is disordered. Inside the valley, as the deepest point is approached from both $x$ or $y$ directions, the height decreases and this means that the equal height contours have diamond-like shape. Using an analysis very similar to that discussed in III B 3, one can explain the clustering of $H$-particles in the IPS phase. We show a typical configuration in Figs. 9(a) and 9(b).

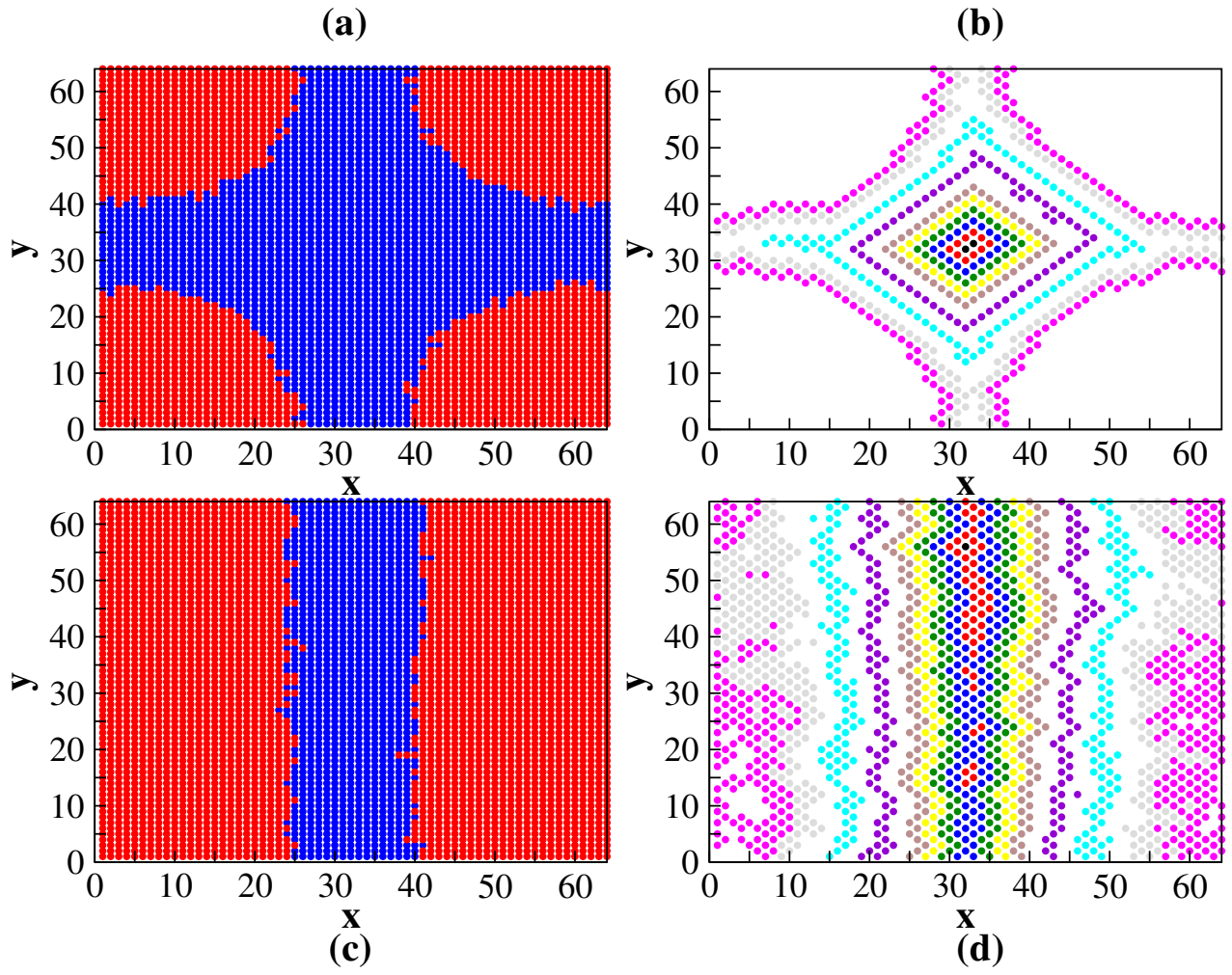

FIG. 9. Representative plots for diamond and wedge type configurations in the IPS phase in two dimensions on a $64 \times 64$ square lattice. In (a) and (c), $H(L)$ clusters are shown in red(blue) while (b) and (d) show the equal-height contour plots.

However, for IPS and FPS phases we have also encountered another type of configuration, where instead of a single point with minimum height, the surface develops a line of minima and the shape of the surface looks like a 'trench' or 'wedge' (see Figs. 9(c) and 9(d)). Through extensive numerical simulations we have also verified that such configurations are finite size effects; for larger systems we only find diamond-shaped contours. Below we estimate the potential energy of particles in a gravitational field for these two difference geometries, and show that the diamond shaped configurations are energetically favorable than the trench-shaped ones. Although the static potential energy may not be the only deciding factor for our nonequilibrium system, nevertheless it gives some indication of the competition between different geometries.

When the equal height contours are diamond shaped, then the number of sites with a given height $z$ above the minimum equals $4z$. For a perfectly ordered configuration, the $H$ particles fill the landscape upto a certain height level $z_0$ above the minimum. Thus the total number of sites occupied by $H$ particles is $1 + 4\sum_{z=1}^{z_0} z$, which must be equal to $\rho N^2$, where $\rho$ is the density of $H$-particles. It follows from this relation that $z_0^2 \approx \rho N^2/2$. Now, the total gravitational energy of the $H$ particles is $\sum_{ij} n_{ij} h(i, j)$, where $n_{ij}$ is the $H$-particle occupancy at site $(i, j)$, and $h(i, j)$ is the height at that site measured from the flat (or logarithmically rough) part of the landscape, in the $L$-region. The number of $H$ particles at a height $z_1$ below the maximum occupied level is $4(z_0 - z_1)$. Hence the total energy becomes

$$E_D = -4 \sum_{z_1=0}^{z_0-1} (z_0 - z_1) z_1 - z_0 = -\frac{2}{3} z_0 (z_0 + 1)(z_0 - 1) - z_0 \approx -\frac{2}{3} \left(\frac{\rho}{2}\right)^{3/2} N^3 \tag{31}$$

to the leading order in $N$.

For a wedge-shaped surface, on the other hand, the equal-height contours are horizontal or vertical lines, running parallel to the line of height minima. The number of sites with a given height $z$ in this case is $2N$ and the highest occupied level $z_0$ in this case is $\rho N/2$. This gives the total energy of $H$-particles in a wedge-like arrangement as

$$E_W = -2N \sum_{z=0}^{\rho N/2} z = -\frac{\rho^2 N^3}{4} - \frac{\rho N^2}{2} \approx -\frac{\rho^2 N^3}{4} \tag{32}$$

to leading order in $N$. For large $N$, it follows from Eqs. 31 and 32 that $E_D < E_W$ unless $\rho$ is very high ($\rho \gtrsim 0.89$). In our simulations, we mainly consider $\rho = 1/2$ and for our case diamond-like arrangements are energetically more favorable for large systems. It will be interesting to study large $\rho$ values to see if wedge-shaped configurations survive even for large $N$. Moreover, we have restricted our studies to a square lattice $N \times N$ here. It may be of interest to see whether different types of arrangements are obtained when a rectangular lattice $N \times M$ or even a different lattice symmetry is considered.

## V. CONCLUSION

In this paper, we have studied different ordered phases present in a coupled nonequilibrium system and explicitly demonstrated how the coupling affects the qualitative nature of the ordering. In our model, a lighter and a heavier particle species move on a potential energy landscape. The particles try to lower the potential energy, and in occupying valleys in the landscape, the heavier species always gets preference over the lighter one. Crucially, the particles also affect the landscape locally, so as to lower the energy further. Depending on how each species interacts with the landscape, we find different phases in the system. In the case when the heavier species tends to push the landscape downward, and the lighter species tends to push it upward, the system shows SPS phase, where the ordering is strongest. When the heavier species pushes the landscape downward, but the lighter one does not push the surface in either direction, rather allows equilibrium local fluctuations of the landscape, we obtain an IPS phase. Finally, the FPS phase is obtained when the lighter species also pushes the landscape downward, but at a smaller rate than the heavier ones. In the limit when both the species affect the landscape in an identical way, either by pushing in a direction with the same rate, or by allowing local equilibrium fluctuations, we obtain FDPO. And in all other cases, we get a disordered phase.

The schematic configurations in Fig. 1 show that the main difference between the SPS, IPS and FPS phases lies in the shape of the landscape. In all these phases, the $H$ and $L$ particles completely phase separate from each other and form one single $H$ and $L$ cluster. But due to the different nature of effects produced by these particles on the landscape, we get these different phases, where the landscape may be completely ordered or may show coexistence of ordered and disordered regions.

In our model, the coupling between the particles and the landscape is such that the mobility of one species depends on the local density of the other. Our linear stability analysis in this case shows that the eigenvalues of the Jacobian matrix that enters the continuity equations are real for $b < -b'$, which implies a homogeneous or disordered state for both particles and landscape. However, for $b > -b'$, the eigenvalues have an imaginary part, which indicates growth of instability which heralds the onset of ordering. It may be worth mentioning here that the nature of cross-species coupling between the mobility and density is crucial. If instead of depending on the density, the mobility of one species depended on higher derivatives of density of the other species, the results might have been different. In [24] a coupled driven system was studied where the mobility of one species depended upon the second derivative of the density of the other species. In that case, however, no ordered phase was found, and homogeneous solutions were shown to remain valid for all parameter regimes.

Finally, it is useful to discuss the implications of our results in the context of some biological systems, in which phase ordering has actually been observed [25]. In a cell membrane, certain species of membrane proteins were found to change the local curvature of the membrane at their binding sites. It has been observed that these proteins form clusters on the membrane, and these clusters in turn generate instability in the membrane shape which gives rise to membrane protrusions. It would be of interest to study these systems within the framework developed in this paper.

## VI. ACKNOWLEDGEMENTS

The computational facility used in this work was provided through the Thematic Unit of Excellence on Computational Materials Science, funded by Nanomission, Department of Science and Technology, India. M.B. acknowledges the award of the J.C. Bose National Fellowship by the Science and Education Research Board, India.